\documentclass{article}
\usepackage{amsmath,amsthm,amssymb}
\usepackage{graphicx}

\newtheorem{proposition}{Proposition}
\theoremstyle{definition}
\newtheorem{definition}{Definition}

\long\def\commented#1{}

\begin{document}

\title{Ensuring Trust in One Time Exchanges:\\ Solving the QoS Problem}
\author{Bernardo A. Huberman, Fang Wu, and Li Zhang\\
\vspace*{3mm} \small{\{bernardo.huberman, fang.wu, l.zhang\}@hp.com}\\
\vspace*{3mm} HP Labs, Palo Alto, CA 94304}

\maketitle

\begin{abstract}
We describe a pricing structure for the provision of IT services
that ensures trust without requiring repeated interactions between
service providers and users.  It does so by offering a pricing
structure that elicits truthful reporting of QoS by providers while
making them profitable.  This mechanism also induces truth-telling
on the part of users reserving the service.

\commented{Experiments with human subjects show that this mechanism
ensures trust in one time exchanges.} \end{abstract}

\pagebreak
\section{Introduction}
\medskip

A most important enabler of interactions over the Internet is the
notion of trust, for it allows access to services, information, and
customers without having to resort to validating procedures that can
increase the complexity of an exchange so as to make it impractical.
That is why reputations play an important role in deciding the level
of trust in commercial exchanges, setting the value of particular
brands, and in deciding whom to consult for professional advice.

A problem with reputations is that the time scales for their buildup
or decay can be very long compared to the typical times involved in
exchanges between a provider and a
user~\cite{hw-reputation-03,rzfk-reputation-00}. This is not a
problem if enough transactions have taken place so that a reputation
or brand name can be established and made known to all parties
concerned. In its place a number of trust management mechanisms for
online environments have been proposed, ranging from the familiar
reputation methods used by eBay, to the creation of trust records
and model based compliance tools for the assurance of utility
computing customers~\cite{bbps-trust-05,jkd-trust-05}.

In all these cases, either history or repeated interactions with a
customer or system are necessary in order to establish a level of
trust. But in many cases, especially those involving one time
exchanges or new providers of services, the repeated interactions
necessary to establish a reputation are not feasible.  Indeed, there
are many situations in online environments where no reputation is
available while needing assurance that a particular quality of
service (QoS) will be provided within a single exchange. An example
could be access to a utility data center that promises a given QoS
level, but the user cannot ascertain whether or not that level of
service will be provided.

In this paper, we consider the most basic model for quality of
service.  In our model, a quality of service contract describes the
likelihood that the service provider delivers the promised service.
We have designed \commented{and tested in the laboratory} a
mechanism that forces the provider to reveal his true assessment of
the probability that he will be delivering a given service in a
single interaction with a user/customer.  We also solved the
complementary truth telling reservation problem of obtaining from
the user his assessment of the true probability that a given level
of resources will be required at the time of their delivery. In both
cases, our mechanisms use a contingent contract to elicit true
revelation of both QoS and likelihood of use through a pricing
structure that forces the parties to make accurate assessments of
their ability to do what they commit to.

We also show that the combination of the two mechanisms, i.e.\ QoS
and truth-telling reservations, provides an approach to solve the
overbooking problem in reservation.  The overbooking problem is when
offering a service, the provider promises delivering more than he is
actually able to do.  The provider's incentive is based on the
belief that some users may not actually use the service due to
uncertainty about their own needs.  However, overbooking can cause
the provider to lose significantly if the user's needs exceeds the
provider's capacity. In our case, by using truth telling
reservations, the service provider can obtain better estimates of
the number of users who actually need the service. We can then
couple those estimates with a quality of service contract to deal
with the overbooking problem.

\paragraph{Related work.} There has been extensive work on the study of  reputation as a basis for  trust~\cite{bbps-trust-05,hw-reputation-03,jkd-trust-05,mrz-peer-05,rzfk-reputation-00}.
 Some recent work~\cite{Jurca2005b,Jurca2003a} also describes  mechanisms that let the users report their past experience  truthfully so as to allow a more accurate buildup of reputations. As  mentioned above, it requires a long time to establish reputations, and  we are primarily interested in interactions for which there is not sufficient past history.
 There are also related papers
 in~\cite{b-trust-01,bs-trust-02} on trading mechanisms for trust  revelation.  They show that one can enforce truthful trust revelation  through the amount of goods to be traded.  In that work the setting is different  from the problem we consider for there is no flexibility  on the amount of goods the buyer purchases.  Instead, we use contingent  contracts to let the service provider reveal the quality of service.

\section{The Quality of Service (QoS) Problem}

\subsection{The problem}

Consider the following one-time exchange problem between two
parties: a \emph{user} and a \emph{service provider}. The user is
interested in buying a service from the provider. Because of
physical constraints, the service provider cannot always provide the
service to a satisfactory level, but can only do so with a
probability $q \in [0,1]$, which we call the \emph{quality of
service (QoS)}. Suppose the service provider knows the real $q$ but
the user does not. We wish to design a mechanism that induces the
service provider to report the true QoS to the user.

\subsection{The mechanism}

We describe a contingency mechanism:
\begin{enumerate}
\item The service provider tells the user his QoS $q'$, which may or may not be the real QoS $q$.
\item The user pays the service provider a \emph{premium} $g(q')$.
\item If the service provider fails to provide the service (to a satisfactory level), he pays the user a \emph{compensation} $h(q')$.
\end{enumerate}
Here $g,h: [0,1]\to \mathbb R^+$ are two functions whose forms are
to be determined.

For a risk-neutral user, he has the following form of expected
utility:
\begin{equation}
\mathbb EU_1 = qv- g(q') + (1-q)h(q'),
\end{equation}
where $v>0$ is his value of using the service, and
$g(q')-(1-q)h(q')$ is his expected cost. Similarly, we assume that
the service provider is risk-neutral, and his expected utility is
given by \begin{equation} \label{eq:provider-utility} \mathbb EU_2 =
g(q')-(1-q) h(q') -c, \end{equation} where $c\ge 0$ is his cost.
Here we have assumed that $c$ is a fixed cost ex ante. That is, the
service provider has to spend this cost whether or not he can
successfully provide the service later. We further assume $c<v$ for
there to be business.

Suppose furthermore that both parties are \emph{rational}, so they
maximize their expected utilities respectively. In particular, the
service provider will report the $q'$ that maximizes
Eq.~(\ref{eq:provider-utility}). Since we wish to induce him to
report the true QoS, Eq.~(\ref{eq:provider-utility}) should be
maximized at $q'=q$. This naturally leads to the following
definition:

\begin{definition}
$(g,h)$ is called \emph{truth-telling} on an interval $I\subseteq
[0,1]$ if for any $q\in I$ and any $q'\in [0,1]$ with $q'\neq q$
\begin{equation} w(q) \equiv g(q)-(1-q)h(q) > g(q')-(1-q)h(q')\,.
\end{equation}
\end{definition}

From this definition, the service provider will report the true QoS
if it is in the truth-telling interval of $(g,h)$. If that happens,
his expected income is $w(q)$. Clearly, he prefers using the
mechanism to not selling any service when $w(q)\ge c$. On the other
hand, knowing that the service provider will report the real $q$,
the user prefers using the mechanism to staying idle when $w(q)\le
v$. We thus make the second definition:

\begin{definition}
$(g,h)$ is called \emph{incentive compatible} on an interval
$I\subseteq [0,1]$ if $c\le w(q)\le qv$ for all $v\in I$.
\end{definition}

To summarize, if $(g,h)$ is truth-telling and incentive compatible
on some interval $I\subseteq [0,1]$, then whenever $q\in I$, the
user and the service provider both want to use the mechanism, and
the service provider reports his real QoS. In the following, we will
give examples of such pairs of $(g,h)$.

\commented{ We summarize what we have shown as a proposition:
\begin{proposition}
\label{prop:qos} Suppose that $(g,h)$ is truth-telling and incentive
compatible on some interval $I\subseteq [0,1]$. If $q\in I$ then the
user and the service provider both want to use the mechanism, and
the service provider reports his real QoS.
\end{proposition}}

\subsection{Realizations}

In this section we give concrete realizations of the functional form
$(g,h)$ that are both truth-telling and incentive compatible.  Such
pairs are not unique.  We give two families of designs with
different forms of compensation.  The most intuitive choice is to
require the compensation proportaional to the quality of service.
Such linear compensation scheme is appropriate for non mission
critical services.

\begin{figure}
\begin{center}
\includegraphics[scale=1.6]{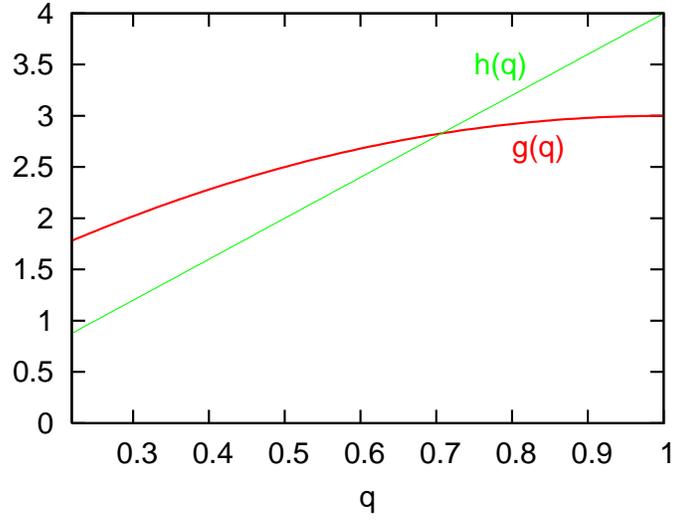}\\(a)
\medskip

\includegraphics[scale=1.6]{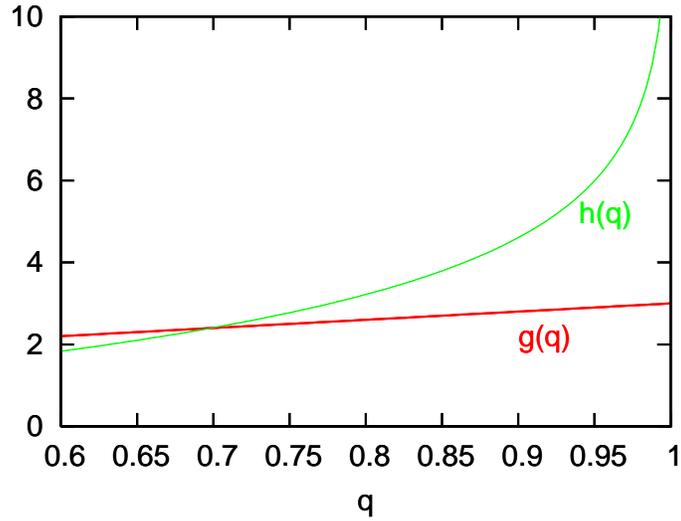}\\(b)
\end{center}
\caption{Plot of the premium and compensation functions for (a)
linear compensation scheme; and (b) logarithmic compensation scheme.
The parameters for both figures are $c=1$, $v=5$, $k=2$, $c_1=1$.
The incentive compatible interval is $[0.219,1]$ for (a) and
$[0.6,1]$ for (b). The curve $g(q)$ is the premium and $h(q)$ is the
compensation.}\label{fig:quad}
\end{figure}

\begin{proposition}
[Linear compensation] Suppose $k,c_1$ are positive numbers
satisfying $c\le c_1 \le v-k$. Let \begin{equation}
q_0=\frac{v-\sqrt{v^2-4kc_1}}{2k}.
\end{equation}
Then the choice
\begin{equation}
g(q)=-kq^2+2kq+c_1, \quad h(q)=2kq
\end{equation}
is truth-telling and incentive compatible on $[q_0,1]$.
\end{proposition}
\begin{proof}
Clearly $g(q), h(q)\ge 0$ for all $q\in [0,1]$. $(g,h)$ is
truth-telling because \begin{equation} g(q')-(1-q)h(q') = -k(q'-q)^2
+kq^2 +c_1 \end{equation} is maximized at $q'=q$ with the value
$w(q)=kq^2+c_1$. The incentive compatibility part is just some basic
algebra.  Figure~\ref{fig:quad}(a) plots the function $g,h$ for a
particular set of parameters.

The parameter $k$ determines the curvature of the function $g$: the
larger $k$, the more curved is $g$.
\end{proof}

The linear compensation scheme is not appropriate for the services
which require high reliability.  In the areas such as in
telecommunications and in financial IT services, the quality of
service is often represented by the ``number of nines'' that
measures how close it is to $1$.  For example ``five nines'' is
equivalent to $q=0.99999$.  For these services it is more
appropriate to set the compensation proportional to $-\log(1-q)$.
This leads to the following design.

\begin{proposition}[Logarithmic compensation] Suppose $k, c_1$ are
positive numbers satisfying $c \le c_1 \le v-k$. Let
$q_0=(c_1+k)/v$. Then the choice
\begin{equation}
g(q)=kq+c_1, \quad h(q)=-k\log (1-q)
\end{equation}
is truth-telling and incentive compatible on $[q_0,1)$.
\end{proposition}
\begin{proof}
Under our assumptions $g(q),h(q)\ge 0$ for $q\in [0,1)$. Consider
the function \begin{equation}
G(q')=g(q')-(1-q)h(q')=kq'+c_1+k(1-q)\log(1-q').
\end{equation}
First order condition
\begin{equation}
G'(q')=k\,\frac{q-q'}{1-q'} =0
\end{equation}
is satisfied when $q'=q$. Since
\begin{equation}
G''(q')=-k\, \frac{1-q}{(1-q')^2}<0,
\end{equation}
$G(q')$ is maximized at $q'=q$ with the maximum value
\begin{equation} w(q) = k(1-q)\log(1-q)+kq+c_1.
\end{equation}
Thus $(g,h)$ is truth-telling on $[0,1)$.

Because
\begin{equation}
w'(q)=-k\log(1-q)\ge 0,
\end{equation}
$w(q)$ is nondecreasing on $[0,1)$. By our assumptions
\begin{eqnarray}
w(0) &=& c_1\ge c,\\
w(1-) &=& c_1+k = q_0v.
\end{eqnarray}
Thus $(g,h)$ is incentive compatible on $[q_0,1)$.
\end{proof}

\paragraph{Remark.}
In both examples, the pairs $(g,h)$ is truth telling in the full
interval $[0,1]$.  It is the incentive compatibility requirement
that restricts the applicable interval.  Therefore, once the service
provider decides to participate, we expect him to be truth-telling.
In both examples, $q_0\to 0$ as $v\to \infty$. In words, the more
valuable the service, the larger is the incentive compatible
interval.

\section{\label{sec:reservation} The reservation problem}

\subsection{The problem}

In the last section we described QoS as the probability that the
service provider can successfully provide the service. The
probability $q$ thus captures the random nature of supply. We now
introduce a second probability $p$ to describe the randomness of
demand. For simplicity, assume that there are two periods: 1 and 2.
With probability $p$  the user will need one unit of service in
period 2, and he knows this probability of needing it in period 1.
The service provider wishes to know the real $p$ in period 1 so that
he can set up the correct quota of service beforehand at a
relatively low cost. In what follows we describe a contingent
contract that incentivizes the user to report his probability of
usage truthfully.

\subsection{The mechanism}

In what follows we will use $p$ and $q$ to denote real
probabilities, and use $p'$ and $q'$ to denote proposed
probabilities.

\begin{enumerate}
\item (Period 1) The service provider tells the user his QoS $q'$, which may or may not be his real QoS $q$.
\item (Period 1) The user tells the service provider a probability $p'$ that he will need the service in period 2, which may or may not be equal to $p$.
\item (Period 1) The user pays the service provider a premium $g(p', q')$.
\item (Period 1) The service provider prepares $p'$ unit of service for the user.
\item (Period 2) If the service provider cannot provide the service (which happens with probability $1-q$), he pays the user a compensation $h(p',q')$. Otherwise, the user can get the service at the price $f(p')$ if he needs it.
\end{enumerate}

Using the mechanism, the user's expected cost is \begin{equation}
\mathbb EC(p',q')=g(p',q')+pqf(p')-(1-q)h(p',q').
\end{equation}
In period 2, the user gets value $v$ if he needs the service and can
obtain it. Hence his value in period 1 can be described by a random
variable \begin{equation} V= \begin{cases} v & \text{with
probability $pq$,}\\ 0 & \text{with probability $1-pq$}.
\end{cases}
\end{equation}
So his expected utility in period 1 is
\begin{equation}
\mathbb EU_1=\mathbb E[V-C]=pqv-\mathbb EC.
\end{equation}

The service provider collects $\mathbb EC$ from each user in
expectation and his cost is $cp$. Thus his expected utility is
\begin{equation} \mathbb EU_2=\mathbb EC-cp.
\end{equation}

If both parties are rational and risk-neutral, then the service
provider maximizes $\mathbb EC(p',q')$ for $q'$ while the user
minimizes $\mathbb EC(p',q')$ for $p'$. So they are playing a
constant sum game.

Similar to the pure QoS problem, we make two definitions:

\begin{definition}
$(g,f,h)$ is called \emph{truth-telling} on a region $I\subseteq
[0,1]^2$ if for any $(p,q)\in I$ and any $(p',q')\in [0,1]^2$ with
$(p',q')\neq (p,q)$ \begin{eqnarray} w(p,q) &\equiv&
g(p,q)+pqf(p,q)-(1-q)h(p,q) \label{eq:w}\\ &>&
g(p',q')+pqf(p',q')-(1-q)h(p',q')\,.
\end{eqnarray}
\end{definition}

\begin{definition}
$(g,f,h)$ is called \emph{incentive compatible} on a region
$I\subseteq [0,1]^2$ if $cp\le w(p,q)\le pqv$ for all $(p,q)\in I$.
\end{definition}

According to the above definition, if $(g,f,h)$ is truth-telling and
incentive compatible on some region $I\subseteq [0,1]^2$, then
whenever $(p,q)\in I$, the user and the service provider both want
to use the mechanism. The user reports his true probability of
usage, and the service provider reports his real QoS. We will show
such triplet $(g,f,h)$ does exist.

\commented{
\begin{proposition}
Suppose that $(g,f,h)$ is truth-telling and incentive compatible on
some region $I\subseteq [0,1]^2$. If $(p,q)\in I$ then the user and
the service provider both want to use the mechanism. The user
reports his true probability of usage, and the service provider
reports his real QoS.
\end{proposition}}

\subsection{Realization}

Following the linear compensation scheme, we can design the
following functions which are truth-telling and incentive
compatible.
\begin{proposition}
Consider the following choice of $(g,f,h)$:
\begin{eqnarray}
g(p',q') &=& k_1p'^2-k_2q'^2+2k_2q'+c_1,\\
f(p') &=& -2k_1p'+c_2,\\
h(p',q') &=& k_1p'^2+2k_2q'+c_3,
\end{eqnarray}
where $k_1, k_2, c_1, c_2, c_3$ are positive numbers satisfying
$c_2\ge 2k_1$, $c_1-c_3\ge c$ and $c<c_1+c_2-k_1+k_2<v$. Then there
exist $p_0<1$ and $q_0<1$ such that $(g,f,h)$ is truth-telling and
incentive compatible on $[p_0,1] \times [q_0,1]$. Furthermore $p_0,
q_0\to 0$ as $v\to \infty$.
\end{proposition}
\begin{proof}
It can be calculated that for our choice of $(g,f,h)$
\begin{equation} \label{eq:ec-tt}\mathbb
EC(p',q')=k_1q(p'-p)^2-k_2(q'-q)^2 -k_1p^2q +k_2q^2 +c_3q +c_2pq
+c_1-c_3,
\end{equation}
so $(p',q')=(p,q)$ is a saddle point of $\mathbb EC(p',q')$. Thus
$(g,f,h)$ is truth-telling on $[0,1]^2$.

From Eq.~(\ref{eq:ec-tt}) we see that
\begin{equation}
w(p,q)= -k_1p^2q+k_2q^2 +c_3q +c_2pq +c_1-c_3.
\end{equation}
By assumption $cp<w(p,q)<vpq$ at $p=q=1$. By continuity, there
exists a region $[p_0,1]\times [q_0,1]$ on which $cp<w(p,q)<vpq$.

Next we show that for any $p_0,q_0>0$ there exists $v$ large enough
such that $(g,f,h)$ is incentive compatible on $I=[p_0,1]\times
[q_0,1]$. Since $I$ is compact, the continuous function $w(p,q)$
achieves its maximum value $w_M$ on $I$. Pick $v$ so large that $v
p_0 q_0 \ge w_M$, and we have $w(p,q) \le w_M \le vp_0q_0 \le vpq$
for $(p,q)\in I$. For the other direction, note that
\begin{equation} \frac{\partial w}{\partial q} = c_2p -k_1p^2 +2k_2q
+ c_3 >0, \end{equation} so $w$ is increasing in $q$. Therefore for
$(p,q)\in I$ we have \begin{equation} w(p,q) \ge -k_1p^2q_0+k_2q_0^2
+c_3q_0 +c_2pq_0 +c_1-c_3 > c_1-c_3 \ge cp.
\end{equation}
Hence $(g,f,h)$ is incentive compatible on $I$.
\end{proof}

\paragraph{Remark.} We note that in the above proof, it is relatively
easier to argue the truth-telling property.  However, it is harder
to prove the incentive compatibility and to calculate the analytical
form of the interval in which the scheme is incentive compatible. It
is possible to design schemes by using the logarithmic compensation
function. It is considerably more challenging for proving the
incentive compatibility.

\section{Application to finite resource reservation}
\medskip

Besides the obvious advantages that our mechanism brings to the
problem of determining QoS in single exchanges and truth-telling in
reservations, there are other scenarios which can also profit from
it. These scenarios are characterized by the fact that resources are
finite and thus they can lead to severe overbooking problems. For
example, consider the case where there are $m$ available units of
resource (plane seats, conference rooms, etc.)  and $n$ users. Each
user may need to consume one unit in period 2.  If $n\le m$ then our
reservation mechanism for infinite resource works without problems.
If $n>m$, however, the service provider can no longer guarantee the
delivery of one unit in period 2, even if the user reserved in
period 1.

This is known as the overbooking problem in a reservation system.
One existing method, used for example in the airline industry, for
dealing with the overbooking problem is by holding an auction at
period 2 until enough users surrender their reservations for
monetary compensation.  This process is often expensive to the
service provider, and can be gamed by some users willing to benefit
from such compensation. Our mechanism can solve this in the
following fashion.

Suppose $n$ users arrive and make reservations one after another. In
period 2 when they claim their needs, they are satisfied one by one
in the order of their arrivals, until there are no more available
resources left. The first $m$ users for sure can be satisfied, so in
period 1 the coordinator simply sells them $m$ plain truth-telling
options with $q=1$ as in the infinite resource problem, and the
users truthfully report their probabilities of needing one unit of
resource in period 2, which we call $p_1, \dots, p_m$.

Now consider the $(m+1)$'th user. The first $m$ probabilities $p_1,
\dots, p_m$ known by the service provider can help him to estimate
the QoS $q_{m+1}$ for the $(m+1)$'th user. For example, under the
assumption that the users' usage probabilities are independent, with
probability $p_1\cdots p_m$ the first $m$ users will each need one
unit of resource in period 2. If that happens, user $m+1$ cannot be
satisfied. Therefore, the probability $q_{m+1}=1-p_1\cdots p_m$ can
be regarded as the QoS for the $(m+1)$'th user. Thus, the
coordinator sells the $(m+1)$'th user an option with QoS $q_{m+1}$,
as specified in Section~\ref{sec:reservation}, and the $(m+1)$'th
user reports his true probability of using one unit of resource in
period 2, denoted by $p_{m+1}$. Once the coordinator knows
$p_{m+1}$, he can calculate the quality of service for the
$(m+2)$'th user, and can sell the $(m+2)$'th user an option with QoS
$q_{m+2}$ and so on. If everyone is risk-neutral and rational, then
all the probabilities $p$ and $q$ will be truthful.

The effectiveness of the procedure just described depends on the
accuracy of the estimation of $q_i$ from $p_1, \dots, p_{i-1}$, a
problem similar to time-series predictions. If the estimation
process is not designed appropriately, the error may accumulate and
propagate to the very last step, i.e.~the service provider will
systematically err in estimating QoS for all subsequent users. In
addition to usage probabilities, the service provider may also take
into account the historical data. \commented{For example, for the
reservation of conference rooms, the provider can set $q_k$ to be
the past statistic that more than $k$ users wanted to use conference
rooms at the same time. But in that case the advantage of a trust
mechanism based on a single exchange is lost.}

Nevertheless in all other cases, including those where the
constraint of finite resources is not severe, we believe that our
contingent contracts can  provide an efficient procedure for
delivering services without resorting to either long histories of
auction systems that can be gamed.

\section{Conclusion}
\medskip

In this paper we presented a mechanism that forces a service
provider to reveal his true assessment of the probability that he
will be delivering a given service in a single interaction with a
user/customer. We also solved the complementary truth telling
reservation problem of obtaining from the user his assessment of the
true probability that a given level of resources will be required at
the time of their delivery. In both cases the mechanism elicits true
revelation of both QoS and likelihood of use through a pricing
structure that forces the parties to make accurate assessments of
their ability to do what they commit to.

Throughout, we considered the most basic model for quality of
service, one in which the service contract describes the likelihood
that the service provider delivers the promised service. We also
showed that the combination of the two mechanisms, i.e.~QoS and
truth-telling reservations, provides an elegant approach to the
overbooking problem in reservation.

There are ongoing experiments with human subjects to verify the
effectiveness of these mechanisms.  The initial results are
encouraging, and the detailed experimental results will soon be
reported.  We also plan to apply the scheme to practical IT service
reservations and to design concrete schemes that are appropriate for
different scenarios.

\bibliographystyle{abbrv}
\bibliography{qos}

\end{document}